\documentclass[lettersize,journal]{IEEEtran}
\usepackage{amsmath,amsfonts}
\usepackage{algorithmic}
\usepackage{algorithm}
\usepackage{array}
\usepackage[caption=false,font=normalsize,labelfont=sf,textfont=sf]{subfig}
\usepackage{textcomp}
\usepackage{stfloats}
\usepackage{url}
\usepackage{verbatim}
\usepackage{graphicx}
\usepackage{cite}
\usepackage{amsmath}
\usepackage{lipsum}
\usepackage{mathtools,amssymb,lipsum}

\usepackage{cuted}
\usepackage{color,soul}
\usepackage{amsmath}
\begin{document}

\title{QoS-based Intelligent multi-connectivity for B5G networks}

\author{Ali Parsa, Neda Moghim , Sachin Shetty,~\IEEEmembership{Senior Member,~IEEE,}
\thanks{This paper was produced by the IEEE Publication Technology Group. They are in Piscataway, NJ.}
\thanks{Manuscript received April 19, 2021; revised August 16, 2021.}
\thanks{(Corresponding author: Neda Moghim) }
\thanks{Ali Parsa is with the Faculty of Computer Engineering, University of Isfahan, Isfaha, Iran (parsa25eng@gmail.com).  }
\thanks{Neda Moghim is with the Faculty of Computer Engineering, University of Isfahan, Isfaha, Iran and Center for Secure and 
Intelligent Critical Systems, Old Dominion University, Norfolk, VA, US (e-mail: n.moghim@eng.ui.ac.ir).  }
\thanks{Sachin Shetty is with the Center for Secure and 
Intelligent Critical Systems, Old Dominion University, Norfolk, VA, US (e-mail: shetty@odu.edu).  }

}

\markboth{Journal of \LaTeX\ Class Files,~Vol.~14, No.~8, August~2021}%
{Shell \MakeLowercase{\textit{et al.}}: A Sample Article Using IEEEtran.cls for IEEE Journals}


\maketitle

\begin{abstract}

The rapid advancement of communication technologies has established cellular networks as the backbone for diverse applications, each with distinct Quality of Service requirements. Meeting these varying demands within a unified infrastructure presents a critical challenge that can be addressed through advanced techniques such as multi-connectivity. Multi-connectivity enables User Equipments to connect to multiple BSs simultaneously, facilitating QoS differentiation and provisioning. This paper proposes a QoS-aware multi-connectivity framework leveraging machine learning to enhance network performance. The approach employs Deep Neural Networks to estimate the achievable QoS metrics of BSs, including data rate, reliability, and latency. These predictions inform the selection of serving clusters and data rate allocation, ensuring that the User Equipment connects to the optimal BSs to meet its QoS needs. Performance evaluations demonstrate that the proposed algorithm significantly enhances Quality of Service (QoS) for applications where traditional and state-of-the-art methods are inadequate. Specifically, the algorithm achieves a QoS success rate of 98\%. Furthermore, it improves spectrum efficiency by 30\% compared to existing multi-connectivity solutions.

\end{abstract}


\section{Introduction}
\IEEEPARstart{A}{s} global connectivity demands rise, 5G-advanced networks are emerging to address the diverse needs of future communication systems. These networks aim to improve data rates, reduce latency, and ensure quality of service (QoS) for applications like ultra-reliable low-latency communications (URLLC) and massive machine-type communications (mMTC)\cite{sufyan20235g}. Multi-connectivity is a promising solution to meet these stringent QoS objectives, enabling simultaneous connections to multiple network access points, thus improving reliability, capacity, and overall user experience.

In cellular networks, multi-connectivity allows the User Equipments (UEs) to connect to multiple access points or radio technologies, enhancing throughput and reliability\cite{sylla2022multi ,schlichter2025evaluating}. Carrier Aggregation (CA), used in LTE and 5G New Radio (NR) systems, combines multiple frequency bands to increase throughput and improve bandwidth utilization\cite{mihovska2020overview}. While CA operates at the physical layer, Dual Connectivity (DC) facilitates connections at higher layers of the protocol stack, allowing for fully separated connections. DC can operate in LTE-NR or NR-only modes, depending on the implementation, offering benefits like high data rates in 5G and LTE coverage. These multi-connectivity techniques ensure consistent, robust connectivity, enhancing applications that require high data transmission and low latency, such as augmented reality, virtual reality, and advanced mobile broadband\cite{agiwal2021survey}.

However, several challenges exist in multi-connectivity, including the need to select cooperative base stations (BSs) (serving cluster selection) \cite{ba2021qos} and efficiently distribute data across multiple access points\cite{jung2021online}. Resource allocation is also a key challenge, as multi-connectivity consumes frequency resources across several BSs, making efficient resource management critical\cite{sylla2022multi}.

Artificial Intelligence (AI) and Machine Learning (ML) can help address these challenges by optimizing resource allocation, enhancing network performance, and improving user experience \cite{10783133}. AI and ML analyze large datasets to predict traffic patterns, dynamically allocate bandwidth, and monitor for anomalies, ensuring better reliability. These technologies can also forecast channel conditions, allowing networks to adapt to changing environments, and support adaptive network policies and energy efficiency, creating more responsive networks that meet the complexities of modern connectivity\cite{juan_jess_hernndezcarln__2022,zia2022ai}.

The shift to Beyond 5G (B5G) networks requires supporting diverse applications with varying Quality of Service (QoS) needs \cite{10820534}. As connected devices grow exponentially, future networks are expected to use higher frequency bands to meet bandwidth demands. However, these bands have poor propagation characteristics, leading to weaker reception and more frequent outages. Multi-connectivity, which allows simultaneous connections to multiple base stations, offers a promising solution by improving reliability and coverage. However, current approaches often neglect or only partially address QoS, leaving opportunities to enhance seamless, application-specific service quality.
This paper presents a QoS-aware multi-connectivity solution that leverages ML to enhance network flexibility and efficiency. Specifically, the proposed algorithm uses Deep Neural Networks (DNNs) to estimate the QoS levels of each BS in terms of data rate, reliability, and latency. These estimations support serving cluster selection and data rate splitting, ensuring optimal connectivity for UEs. 

The primary contributions of this paper are summarized as follows:
\begin{itemize}
    \item \textbf{Intelligent Serving Cluster Selection:} The proposed algorithm dynamically connects the UE to the optimal set of BSs, rigorously ensuring compliance with stringent QoS demands. Unlike conventional methods, this selection is driven by real-time QoS metrics, maximizing service continuity and user experience.
    \item \textbf{Adaptive Data Rate Splitting:} The UE’s total traffic is optimally distributed across the selected BSs, guaranteeing precise adherence to critical QoS parameters—data rate, ultra-low latency, and high reliability. This fine-grained allocation enhances spectral efficiency while minimizing resource wastage.
    \item \textbf{Holistic QoS-Aware Multi-Connectivity:} By jointly optimizing data rate, reliability, and latency in a unified multi-connectivity framework, the proposed algorithm significantly enhances QoS for applications where existing methods fall short. It achieves a 98\% QoS success rate and improves spectrum efficiency by 30\%, outperforming traditional approaches that treat these metrics in isolation.
\end{itemize}

Performance evaluation results demonstrate significant improvements in QoS for applications where standard methods and state-of-the-art algorithms fall short. Additionally, the proposed algorithm enhances spectrum efficiency compared to existing multi-connectivity solutions. 

The proposed algorithm can enhance QoS satification level for various consumer electronics devices and services. For instance, the majority part of applications that run on the smartphones and tablets require high data rates for video streaming while wearables demand low latency for real-time health monitoring, and smart home systems prioritize reliable connections for automation and security. By dynamically selecting the optimal serving cluster, the proposed algorithm effectively meets these diverse QoS requirements, improving the overall user experience in B5G networks.

This paper is organized as follows: Section II reviews related works, providing context and identifying gaps in existing research. Section III presents the proposed algorithm, detailing its design and innovative techniques. Section IV offers a comprehensive performance evaluation that assesses the effectiveness of the algorithm using metrics such as data rate, latency, and reliability. Finally, Section V concludes the paper by summarizing key findings and discussing implications for future research.
\section{Related Works}
As previously mentioned, several challenges are associated with multi-connectivity. Previous research has contributed to addressing these challenges, and this section outlines some of them while examining the research conducted.

One significant challenge is determining the optimal number of gNBs that should simultaneously serve a UE, known as the degree of multi-connectivity. Reference \cite{gapeyenko2018degree} presents a method for selecting the degree of multi-connectivity by balancing outage probability and system complexity. While a higher degree of multi-connectivity reduces outage probability, it also increases system overhead due to the more frequent state information updates required between the UE and BSs. Moreover, a higher degree of multi-connectivity demands additional frequency resources from BSs, affecting spectrum efficiency. Results suggest that increasing the degree beyond four provides negligible improvements in outage probability reduction. 
Similar to \cite{gapeyenko2018degree}, our proposed algorithm determines the degree of multi-connectivity. However, in addition to this, it explicitly identifies the members of UE's serving cluster.


Reference \cite{mahmood2019resource} investigates the resource consumption associated with multi-connectivity. The study proposes a model to quantify outage probability and resource usage in multi-connectivity scenarios, comparing it to single-connectivity. The results show that while multi-connectivity significantly reduces outage probability, it increases resource consumption, suggesting its use in applications requiring high reliability.
Building on the findings of \cite{mahmood2019resource}, which highlight the increased resource consumption of multi-connectivity, we consider spectrum efficiency as a key parameter in the decision-making process of our proposed algorithm.

Reference \cite{wolf2018rate} examines the trade-off between rate and reliability in multi-connectivity. An analytical framework is presented to calculate outage probability and system throughput, incorporating factors such as the number of connections, modulation scheme, bandwidth, coding rate, and signal-to-noise ratio. The framework offers two plans: one prioritizing outage probability and the other maximizing system throughput.
In line with this, our proposed algorithm adopts a comprehensive approach to QoS by balancing key parameters, including data rate, reliability, and latency, to achieve an optimal trade-off.

In \cite{sana2019multi}, sum-rate optimization is formulated as a non-convex problem and solved using a multi-agent RL framework, where each UE selects its serving cluster. In contrast, our heuristic-based approach enhances scalability by incorporating latency considerations and efficiently determining both serving cluster members and their participation factors.

In \cite{weedage2023impact}, the impact of multi-connectivity on outage probability and throughput is evaluated using a stochastic geometry model. The results show that while multi-connectivity significantly reduces outage probability, it does not necessarily increase the data rate for all users. Based on these findings, the study suggests that multi-connectivity should be selectively applied, particularly for users at the cell edge, with mechanisms in place to ensure fairness.

In addition to selecting a serving cluster, the participation of serving cluster members in data transmission must be considered. Reference \cite{hernandez2022deep} proposes a method for data splitting in multi-connectivity, where each gNB’s participation is determined by the required data rate. The study uses a Deep Reinforcement Learning (DRL) agent to balance load and avoid overloading gNBs while ensuring the overall data rate. This research demonstrates improved communication throughput, with an expanded version published in \cite{hernandez2022deep1}.

Similar to \cite{hernandez2022deep1}, our proposed algorithm optimizes multi-connectivity by selecting serving cluster members and their participation factors. Key differences include incorporating latency considerations, which \cite{hernandez2022deep1} does not explicitly address. Additionally, while \cite{hernandez2022deep1} employs DRL, leading to scalability challenges as gNBs increase, this paper proposes a heuristic approach for efficient and scalable multi-connectivity configuration.

\section{Proposed Algorithm}
The primary objective of the proposed algorithm is to provide a QoS-based multi-connectivity solution. Specifically, the method aims to determine the gNBs to which a UE should connect and how the UE’s traffic should be distributed to meet the desired QoS levels. To achieve this, two main challenges must be addressed:
\begin{itemize}
  \item Serving Cluster Selection: With the smaller cell sizes in 5G and beyond, the number of gNBs has increased compared to previous generations, providing multiple options for connecting a UE. A serving cluster is a subset of gNBs that simultaneously serve a specific UE. The cluster must be selected to maximize the ability to meet the UE’s QoS requirements.
  \item Determining the Participation Share of each gNB: After selecting the serving cluster, each gNB must contribute to meeting the UE’s QoS requirements. However, due to variations in the UE's reception quality and each gNB's available resources, their participation rates may vary. It is essential to determine the portion of traffic that each gNB should handle to ensure optimal performance.
\end{itemize}
\subsection{System Model}
\begin{figure}[!t]
\centering
\includegraphics[width=3.4in]{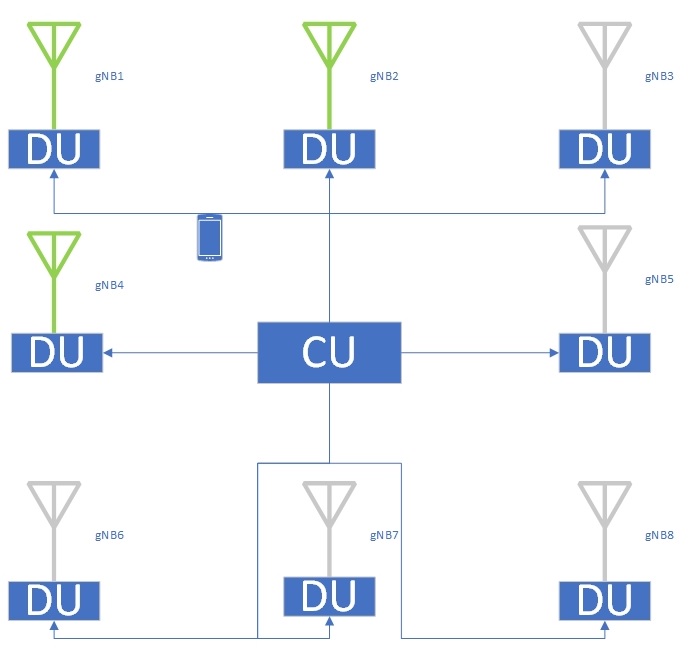}
\caption{Network Architecture of the System Model}
\label{fig_11}
\end{figure}

We considered a scenario with several gNBs in a 5G network using O-RAN architecture. A subset of these gNBs forms the serving cluster to transmit data to the UE. Each gNB connects to a decentralized unit (DU), which handles the RLC, MAC, and PHY layers based on functional split. All DUs are linked to a central unit (CU) that implements the higher layers. Fig. \ref{fig_11} illustrates this setup. 

Connecting UEs to multiple gNBs simultaneously can significantly impact their QoS. Therefore, before introducing the proposed method, we will analyze these effects in detail. To achieve this, an optimization problem is formulated, with the variables used outlined in Table \ref{tab:table1}. 
\begin{table*}[!t]
\caption{The notation used for equations\label{tab:table1}}
\centering
\begin{tabular}{|w{c}{0.1\linewidth} | w{c}{0.2\linewidth}| p{0.7\linewidth}|}
\hline
\textbf{Notation} & \textbf{Variable Type and Dimension} & \textbf{Description}\\
\hline
\textit{B} & Integer & The total number of gNBs in the network\\
\hline
\textit{CON} & Binary vector \textit{  $ 1\cdot B $ } & Connectivity vector. If the i-th component is equal to one, it indicates an active connection between the UE and the gNB with index i. The serving cluster for the UE can be derived from this vector \\
\hline
$\mu$ & Integer Vector \textit{ 1 $\cdot$ B} & The i-th component represents the OFDM numerology of the gNB with the corresponding index. \\
\hline
$NS$ & Integer & The number of time slots for each gNB. It can be computed from $\mu$. \\
\hline
$MCS$ & Integer Matrix $B \cdot NS $ & The MCS used for the connection between the UE and gNB in each time slot. Due to the variation in the number of time slots caused by differences in the OFDM numerology parameters across different gNBs, the maximum number of definable time slots is considered as the second dimension of the matrix. The values in this matrix range from 1 to 27. \\
\hline
$BPMS$ & Integer Vector \textit{ 1 $\cdot$ 28} & The number of bits in each modulation symbol. Given that there are 28 different options for the modulation and coding schemes, this variable is represented by a 28-element vector, with values of 2, 4, 6, and 8. \\
\hline
$CR$ &  Integer Vector \textit{ 1 $\cdot$ 28}  & The coding rate of each modulation and coding scheme \\
\hline
$PWRLV$ & Integer Vector \textit{ 1 $\cdot$ B} & The power level used per gNB. It is assumed that there are 3 power levels available.\\
\hline
$Rate$ & Integer & The experienced data rate of UE. \\
\hline
$RateReq$ & Integer & The data rate requirement of the UE. \\
\hline
$RelReq$ & Decimal & The reliability requirement of the UE. \\
\hline
$LatReq$ & Decimal & The latency requirement of the UE. \\
\hline
$MaxLat$ & Decimal Vector $1 \cdot B $& The maximum delay for each gNB \\
\hline
$CQIHist$ & Integer Matrix \textit{ B $\cdot$ CHL} & The matrix of the last state of CQIs received from the UE in each of the gNBs \\
\hline
$CHL$ & Integer & The length of the CQI History \\
\hline
$MaxClSize$ & Integer & The maximum size of the serving cluster \\
\hline
\end{tabular}
\end{table*}

The optimization problem aims to maximize spectrum efficiency. It assumes a downlink data transmission to the UE over a duration of $t$ seconds. Spectrum efficiency, in this context, is calculated as the ratio of the data rate experienced by the UE to the frequency resources allocated for the transmission, as expressed in Equation \ref{eq1}.
\begin{equation}
\label{eq1}
SE = \frac{DR}{CBW}
\end{equation}
In Equation \ref{eq1}, $DR$ denotes the data rate experienced by the UE, and $CBW$ represents the consumed frequency resources. The data rate for the UE can be determined using Equation \ref{eq2}.
\begin{equation}
\label{eq2}
DR = \sum_{i=1}^{B} DR_i
\end{equation}
Based on Equation \ref{eq2}, the total data rate experienced by the UE is the sum of the rates experienced from each member of its serving cluster. The rate experienced from each gNB can be calculated using Equation \ref{eq3}:
\begin{equation}
\label{eq3}
DR_i = \frac{ \sum_{j=1}^{NS_i}  (1-BLER_{i,j}) \times TBSD_{i,j} }{t}
\end{equation}
In Equation \ref{eq3}, ${BLER}_{i,j}$ represents the block error rate of the connection between the UE and the gNB with index $i$ in time slot $j$. Additionally, $TBDS_{i,j}$ indicates the data size in the transport block for the connection between the UE and the gNB with index $i$ in time slot $j$. The variable $NS_i$ in Equation \ref{eq3} represents the number of time slots in the time interval of $t$ seconds, and it can be calculated using Equation \ref{eq4}.
\begin{equation}
\label{eq4}
NS_i = t \times 2 ^ {\mu_i} \times 10^3
\end{equation}
In Equation \ref{eq4}, $\mu_i$ represents the OFDM numerology of the connection between the UE and the gNB with index $i$. The data size of the transport block is a function of the number of frequency resources assigned to the UE, as well as the modulation and coding scheme. It can be calculated using Equation \ref{eq5}.

\begin{equation}
\label{eq5}
TBSD_{i,j} = 156 \times ALRB_i \times CR_{MCS_{i,j}} \times BPMS_{MCS_{i,j}}
\end{equation}
In Equation \ref{eq5}, the coefficient 156 shows the maximum number of Resource Elements (REs) that can be used for data transmission in each Resource Block (RB). $ALRB_i$ indicates the allocated RBs to the UE on the gNB with index $i$.
Also, $BLER_{i,j}$ is a function of the channel status between the UE and gNB with index $i$, modulation and coding scheme, and transmission power, but there is no closed-form relationship to calculate it. Therefore, ML methods will be used for estimating it.
The amount of resources used in the period $t$ is given by the sum of the resources used by each member of the serving cluster, as indicated in the denominator of Equation \ref{eq1}. Therefore, the amount of frequency resources used on a gNB for the intended UE can be calculated using Equation \ref{eq6}.
\begin{equation}
\label{eq6}
CBW_i = NS_i \times ALRB_i \times 12 \times 15 \times 10^3
\end{equation}

In this way, the total amount of used resources is calculated using Equation \ref{eq7}.
\begin{equation}
\label{eq7}
\begin{aligned}
CBW & = \sum_{i=1}^{B} CBW_i \\
&=\sum_{i=1}^{B} CON_i \times NS_i \times ALRB_i \times 12 \times \\ & 15 \times 2^{\mu_i} \times 10^3 \\
&=\sum_{i=1}^{B} CON_i \times t \times 2^{\mu_i} \times ALRB_i \times 12 \times \\ & 15 \times 2^{\mu_i} \times 10^6
\end{aligned}
\end{equation}

After calculating the throughput, delay is another important parameter. From the perspective of the MAC layer, latency can be affected by the length of the time slot and the re-transmissions of the HARQ process. In the best case, the data sent from the gNB side will be decoded in the initial transmission by the UE, while in the worst case, the decoding process will occur after 3 re-transmissions. Therefore, the minimum and maximum latency can be calculated using Equations \ref{eq8} and \ref{eq9}, respectively.
\begin{equation}
\label{eq8}
MinLat_i = \frac{2}{2^{\mu_i}} \times 10^{-3}
\end{equation}

\begin{equation}
\label{eq9}
MaxLat_i = \frac{8}{2^{\mu_i}} \times 10^{-3}
\end{equation}

Reliability is the third QoS requirement that must be considered. From the physical layer perspective, the block error rate (BLER) is the most important parameter related to reliability. Due to the HARQ mechanism, some physical layer errors can be corrected, meaning the reliability from the physical layer perspective is always less than or equal to the reliability from the MAC layer perspective. Since there is no closed-form relationship to calculate the error rate, ML methods are used to estimate this parameter. After defining the relationships among the various parameters, the optimization problem can be presented in Equations 10 to 14.

To solve the optimization problem, it is important to note that there is no closed-form relationship for the variables $BLER_{i,j}$ and $Rel_i$. Therefore, ML-based methods are employed to address this issue. The next subsection will propose a solution to this problem.
\begin{equation}
\label{eq10}
\begin{aligned}
 \max_{CON,MCS,PWRLV} \hspace{2cm} \\ 
\frac{1}{t^2} \times \sum_{i=1}^{B} [ \frac{1}{CON_i \times 
        ALRB_i \times 
        180 \times 
        2^{\mu_i} \times 
        10^6}    
        \times
        \\
        \sum_{j=1}^{\mu_i} 
        [(1- BLER_{i,j}) \times
        156 \times
        ALRB_i \times \\
        CR_{MCS_{i,j}} \times 
        BPM_{MCS_{i,j}} ]
        ] 
\end{aligned}
\end{equation}
$ S.t: $
\begin{equation}
\label{eq11}
DR \geq RateReq
\end{equation}
\begin{equation}
\label{eq12}
MinLat_i  \leq LatReq
\end{equation}
\begin{equation}
\label{eq13}
 Rel_i \geq  RelReq
\end{equation}
\begin{equation}
\label{eq14}
 \sum_{i=1}^{B} CON_i \geq 1
\end{equation}

\subsection{BLER Estimation}
In the proposed method, the BLER is considered a function of transmission power, modulation and coding scheme, sub-carrier spacing, and channel status. This section focuses on estimating the BLER as a function of these variables using ML methods, which will then be integrated into the serving cluster selection.

\begin{figure}[!t]
\centering
\includegraphics[width=2.5in]{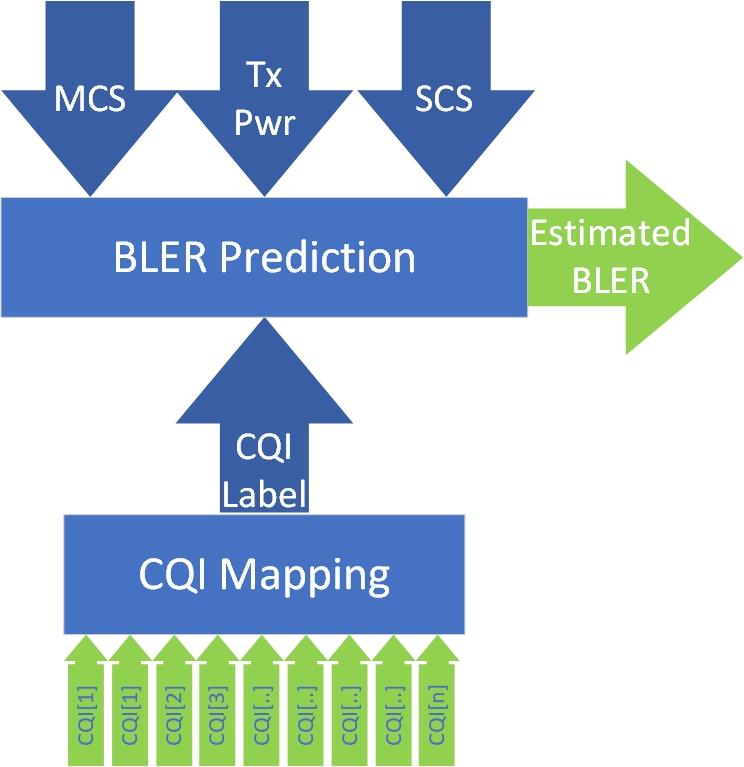}
\caption{The BLER estimation model}
\label{fig_12}
\end{figure}

All parameters influencing the BLER estimation are mostly categorical in nature. However, the channel status, represented by a sequence of CQI values, is a high-dimensional parameter that cannot be used in its raw form. Therefore, a space transformation process is applied to map the CQI sequence to a single value. This mapped value is then used for the BLER estimation. Fig. \ref{fig_12} shows a general view of this process.

In the proposed algorithm, a sequence of CQI values is used instead of a single value to address issues like discretization errors. To process this sequence, it is mapped to a single value (label) using a Long-Short-Term Memory (LSTM) model. The CQI sequence, which has an n-dimensional space, is the input to the LSTM model, and the output is the corresponding CQI label.

\begin{figure}[!t]
\centering
\includegraphics[width=3.4in]{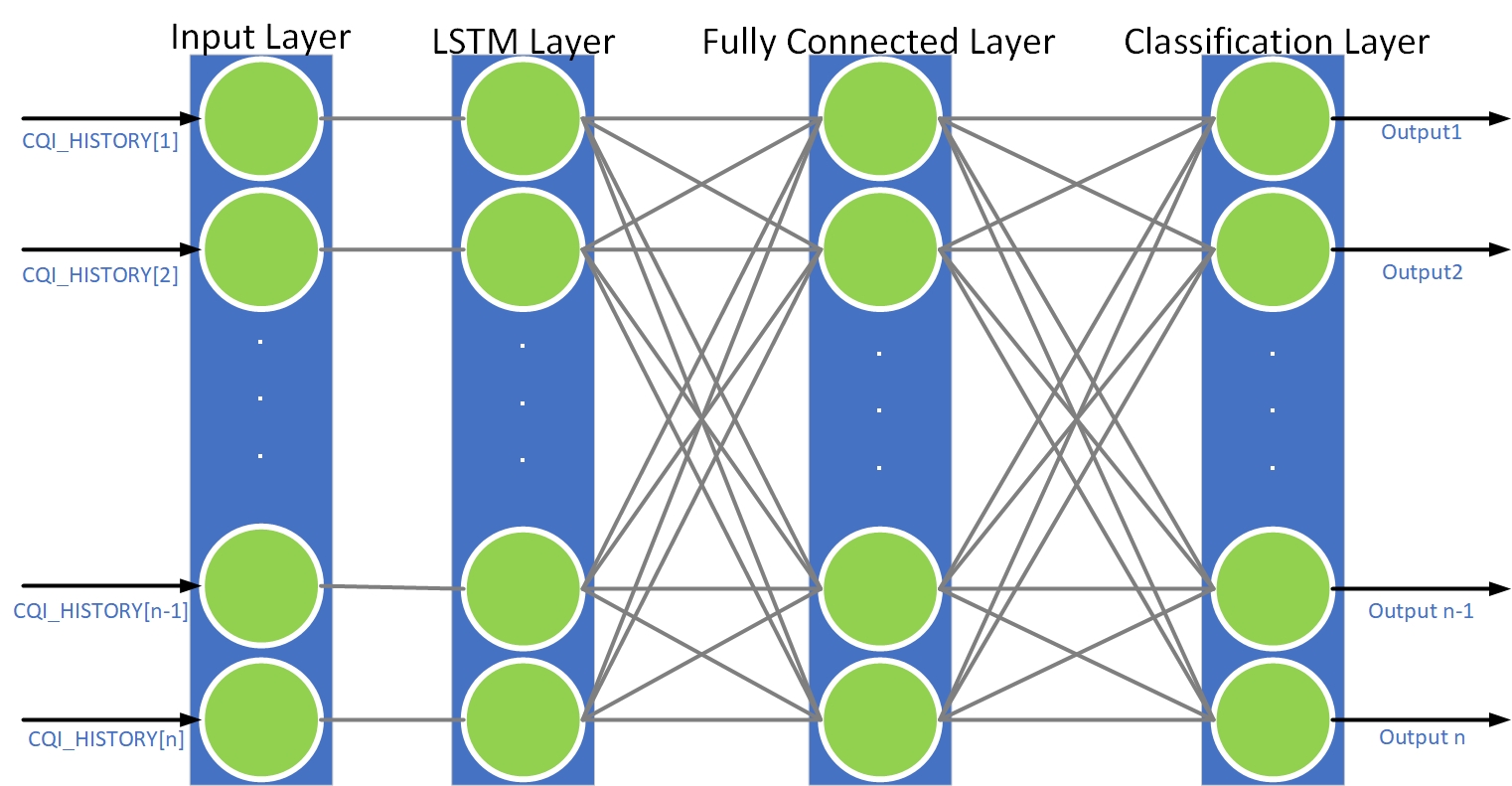}
\caption{The LSTM model for estimating the CQI label}
\label{fig_13}
\end{figure}

\begin{figure}[!t]
\centering
\includegraphics[width=3.4in]{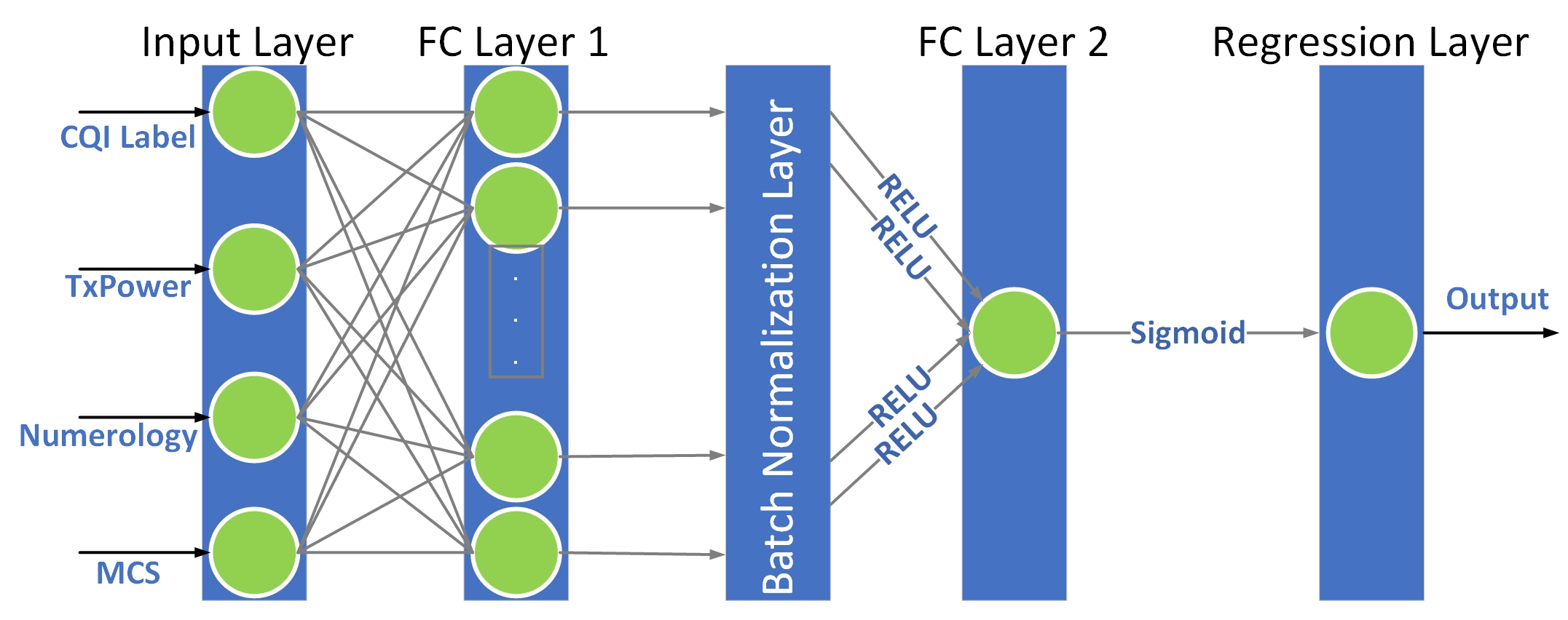}
\caption{A view of the block error rate estimation neural network}
\label{fig_14}
\end{figure}

Once the CQI sequence is mapped to a CQI label, BLER estimation begins. This estimation uses multidimensional regression, with the CQI label, sub-carrier spacing, transmission power, and modulation and coding scheme as inputs. Regression is performed using a DNN. In the penultimate layer of this DNN, a sigmoid activation function is applied to constrain the output values within the range of zero to one, which is the acceptable range for the BLER. Figs \ref{fig_13} and \ref{fig_14} illustrate the mapping and regression stages, respectively.

\subsection{Serving cluster selection and participation rate}
After explaining BLER calculation, the process for selecting serving clusters and their roles in UE data transmission is outlined. The problem involves two main steps: selecting a subset of gNBs to serve the UE and determining each gNB's participation rate. A multi-step method is used to address this.

\begin{itemize}
    \item \textbf{Step 1:} The block error rate (BLER) for each gNB is estimated across all modulation and coding schemes using the presented model. This results in a matrix called $gNBsBLER$, where $gNBsBLER_{i,j}$ represents the BLER for using the $j$-th modulation and coding scheme at the $i$-th gNB.
    \item \textbf{Step 2:} The minimum latency (Equation \ref{eq8}), estimated maximum data rate, and spectrum efficiency (Equation \ref{eq1}) for each gNB are calculated. The estimated rate is determined using Equation \ref{eq101}, which simplifies Equation \ref{eq3}.
    \begin{equation}
    \label{eq101}
    ESRate_{i,j}=\frac{(1-BLER_{i,j}) \times NS_i \times TBSD_{i,j}}{t}
    \end{equation}
    In Equation \ref{eq101}, the variable $gNBsBLER_{i,j}$ comes from Step 1, while $NS_i$ and $TBDS_{i,j}$ are derived from Equations \ref{eq4} and \ref{eq5}, respectively. The maximum element in each row of the ${ESRate}_{i,j}$ matrix represents the maximum estimated rate for the $i$-th gNB and is denoted as $MAXESRate_i$. The index corresponding to this maximum is represented as ${MCS}^*$. Similarly, the maximum spectrum efficiency achievable for a gNB is calculated using Equation \ref{eq111}.
    \begin{equation}
    \label{eq111}
        ESSE_i=\frac{MAXESRate_i}{CBW_i}
    \end{equation}
    \item \textbf{Step 3:} Using the results from the previous step and the UE's QoS requirements, separate scores for data rate, latency, reliability, and spectrum efficiency are computed. These scores are determined using Equations \ref{eq112} to \ref{eq115}.
     \begin{equation}
    \label{eq112}
        RateScore_i=min (1,\frac{MAXESRate_i}{RateReq})
    \end{equation}

    \begin{equation}
    \label{eq113}
        RelScore_i=min (1,\frac{gNBsBLER_{i,MCS^*}}{RelReq})
    \end{equation}

    \begin{equation}
    \label{eq114}
        LatScore_i=min (1,\frac{LatReq}{MinLat_i})
    \end{equation}

    \begin{equation}
    \label{eq115}
        ESScore_i=\frac{ESSE_i}{max(ESSE)}
    \end{equation}

    \item \textbf{Step 4:} The overall score for each gNB is derived from the four scores calculated in the previous step. This score is determined using Equation \ref{eq16}.
    \begin{equation}
    \label{eq16}
    \begin{aligned}
        OS_i= \alpha \times RateScore_i + \beta \times RelScore_i + \\ \gamma \times LatScore_i + \delta \times ESScore_i
    \end{aligned}
    \end{equation}
    As shown in Equation \ref{eq16}, the overall score for each gNB is calculated by linearizing the scores related to the three QoS characteristics and the spectrum efficiency score. The coefficients assigned to each score allow flexibility in prioritizing specific QoS characteristics alongside spectrum efficiency.
    
    \item \textbf{Step 5:} The gNBs are sorted based on the scores from the previous step. An empty set is initialized as the serving cluster (SC). gNBs are then added to the SC set according to the sorted list. This process continues until either the total estimated rate of the gNBs in the SC meets or exceeds the UE's required rate, or the SC reaches its maximum size.

    \item \textbf{Step 6:} The participation rate of each gNB is determined based on its maximum estimated data rate. These ratios are calculated using Equation \ref{eq17}.
    \begin{equation}
    \label{eq17}
        CF_i=\frac{MAXESRate_i}{\sum_{k \leq |SC|} MAXESRate_k}
    \end{equation}

    In Equation \ref{eq17}, the $CF_i$ variable represents the participation coefficient of the $i$-th member in the serving cluster set during the data transfer process.
\end{itemize} 

\section{Performance Evaluation}
In this section, the efficiency of the proposed algorithm is assessed through an extensive simulation study. First, the performance of the BLER estimation model is evaluated in Subsection A. Then a comparison of the proposed algorithm's efficiency against benchmark algorithms will be presented in Subsection B.

\subsection{BLER Estimation Model}
\textbf{Training Process of the LSTM network for CQI mapping}

To train the CQI mapping LSTM network, labeled data is required. A training dataset is generated using a simulator, ensuring sufficient variety in the data by considering different sub-carrier spacings, transmission power levels, and modulation and coding schemes. Additionally, 10 random locations are selected for each parameter set to enhance data variety.

\begin{figure}[!t]
\centering
\includegraphics[width=3.4in]{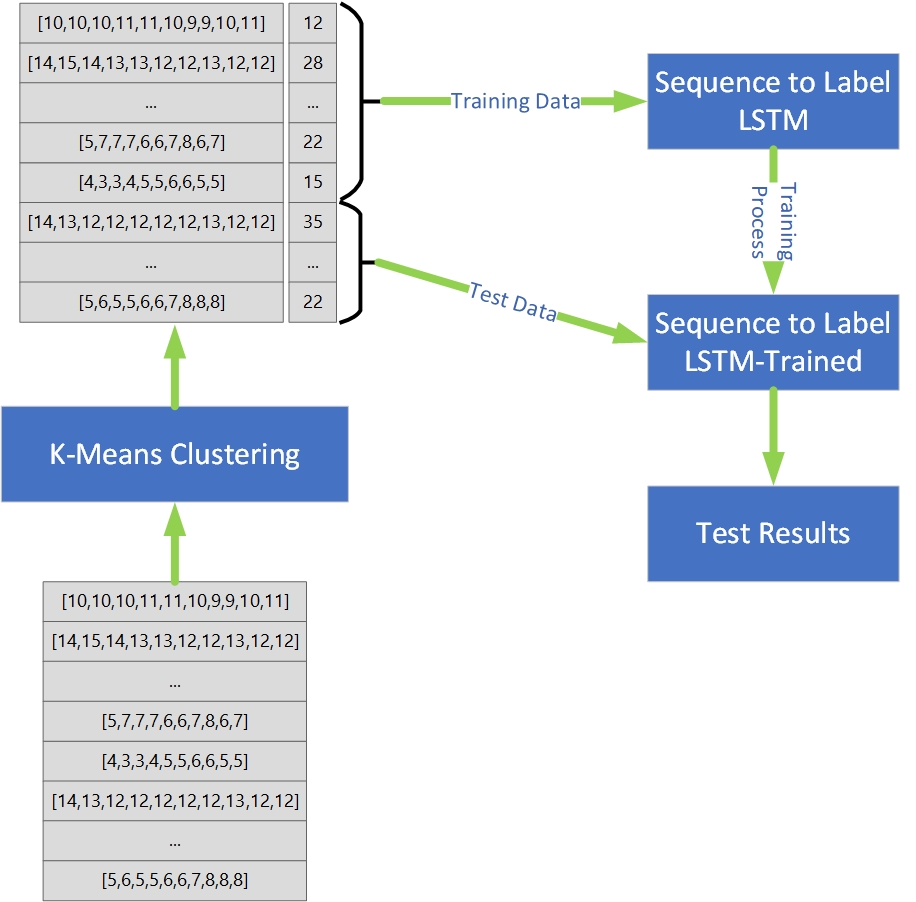}
\caption{Training process of the LSTM network for CQI mapping}
\label{fig_15}
\end{figure}

CQI data is initially unlabeled, so a clustering operation is applied using the k-means algorithm. This divides the data into clusters, and the cluster ID serves as the label for the corresponding CQI sequence. The training process uses these labeled data. As shown in Figure \ref{fig_15}, after clustering, the data is divided into clusters, with each cluster labeled by its ID. The data is then split into 70 percent for training and 30 percent for testing. To improve training reliability, the data is shuffled for each epoch. Table \ref{tab:table2} outlines the parameters used in the LSTM network. 
It is important to note that the LSTM network's hyperparameters and architecture were optimized through iterative trial and error, training the model with various configurations. The final results correspond to the parameter set that achieved the highest accuracy, as shown in Table II.

\begin{table}[!t]
\caption{Parameters used in LSTM network\label{tab:table2}}
\centering
\begin{tabular}{|c||c|}
\hline
\textbf{Parameter} & \textbf{Value}\\
\hline
Gradient Threshold & 1\\
\hline
Maximum Training Epochs  & 50 \\
\hline
Batch size & 128 \\
\hline
LSTM Type & Bi-LSTM \\
\hline
No.of Neurons in input Layer & 50 \\
\hline
No.of Neurons in LSTM Layer & 100 \\
\hline
No.of Neurons in FC Layer & 40 \\
\hline
No.of Neurons in output Layer & 40 \\
\hline
Activation Function (Only before output) & Softmax \\
\hline
Optimizer & ADAM \\
\hline
Shuffling Strategy & every epoch \\
\hline
\end{tabular}
\end{table}

\begin{figure}[!t]
\centering
\includegraphics[width=3.4in]{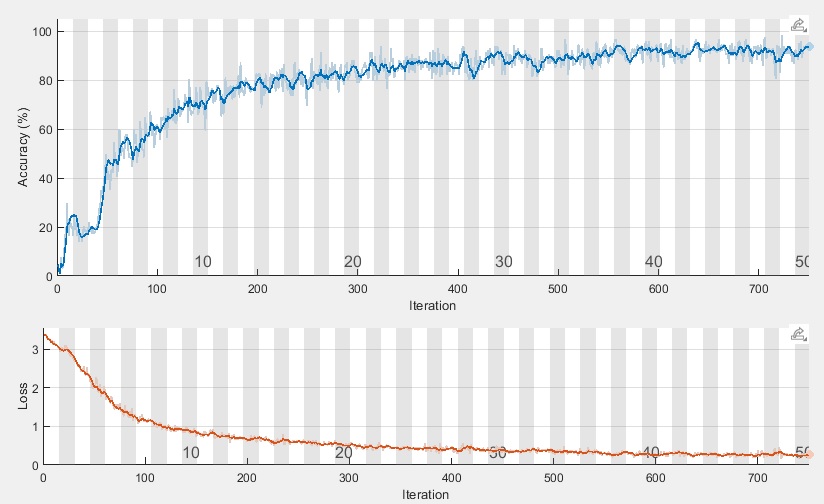}
\caption{accuracy and loss during the training process of CQI mapping LSTM network}
\label{fig_16}
\end{figure}

Figure \ref{fig_16} illustrates the accuracy and loss results during the training process. As shown, the LSTM network progressively learns the labels associated with the CQI sequences, achieving an accuracy of over 90 percents by the end of the training.

\begin{figure}[!t]
\centering
\includegraphics[width=3.4in]{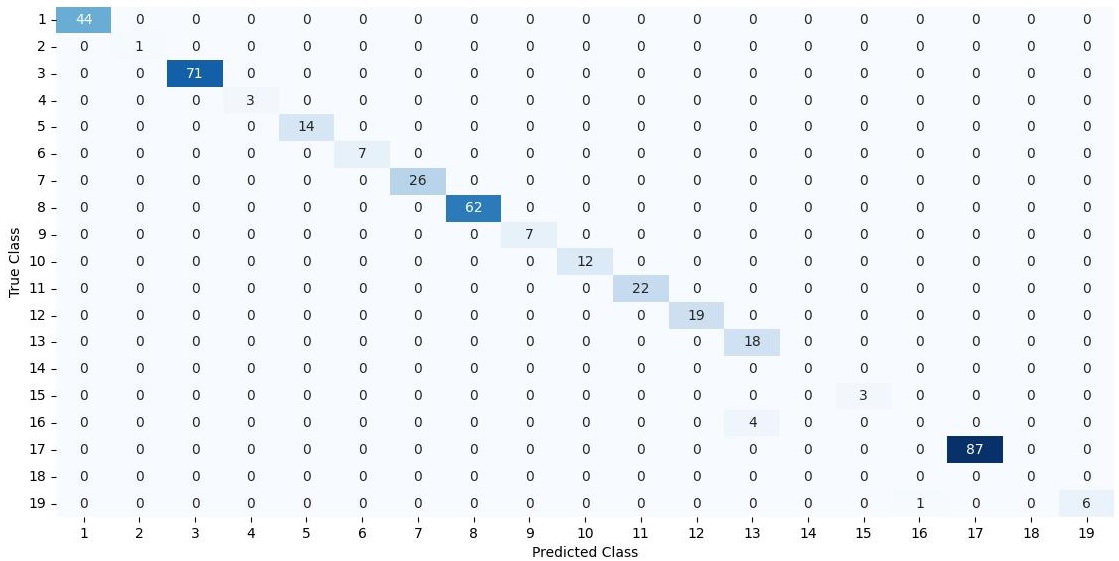}
\caption{Confusion matrix for 20 classes}
\label{fig_17}
\end{figure}

\begin{figure}[!t]
\centering
\includegraphics[width=3.4in]{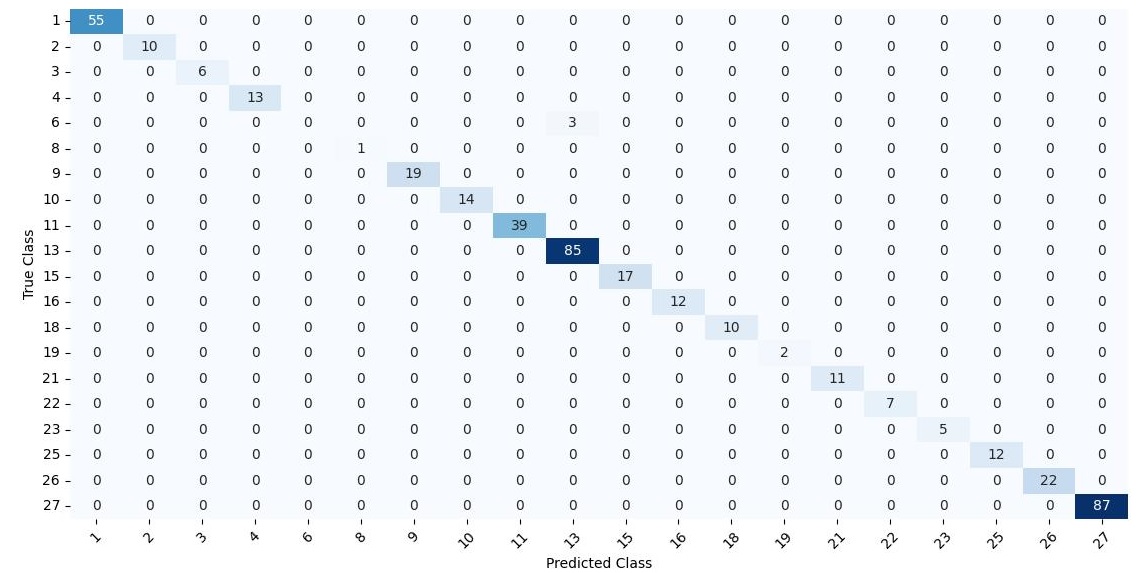}
\caption{Confusion matrix for 30 classes}
\label{fig_18}
\end{figure}

\begin{figure}[!t]
\centering
\includegraphics[width=3.4in]{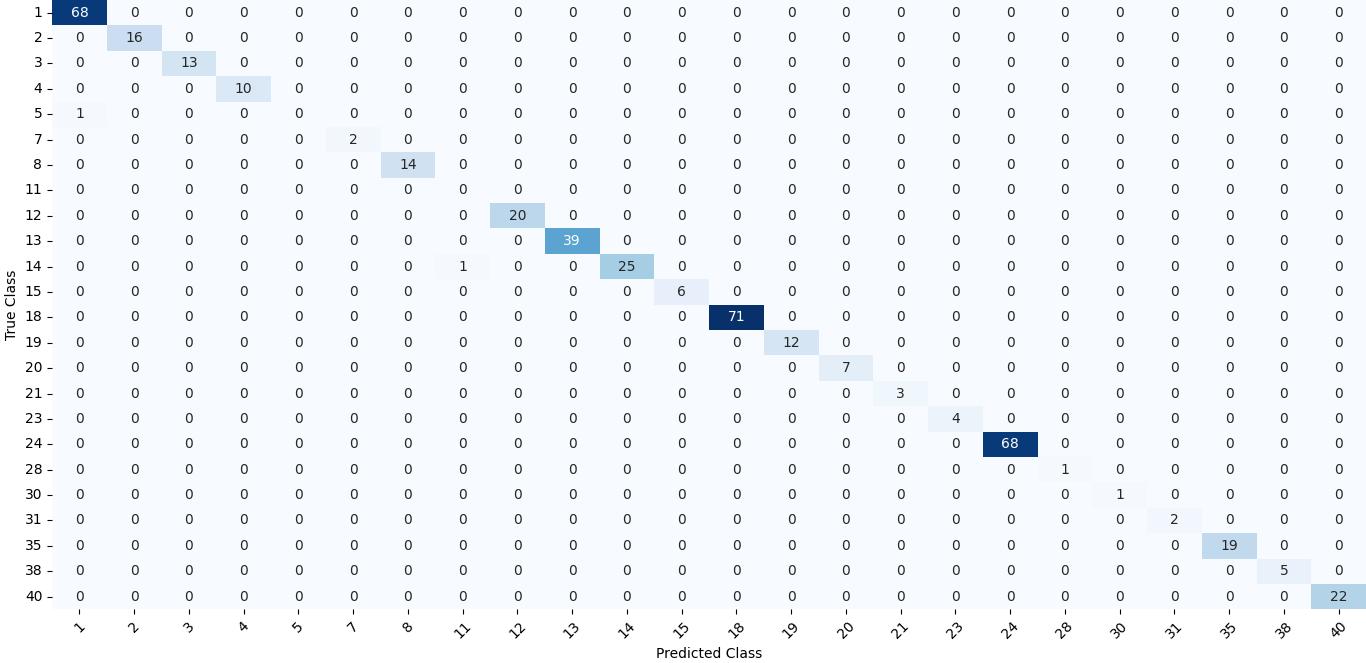}
\caption{Confusion matrix for 40 classes}
\label{fig_19}
\end{figure}

\begin{figure}[!t]
\centering
\includegraphics[width=3.4in]{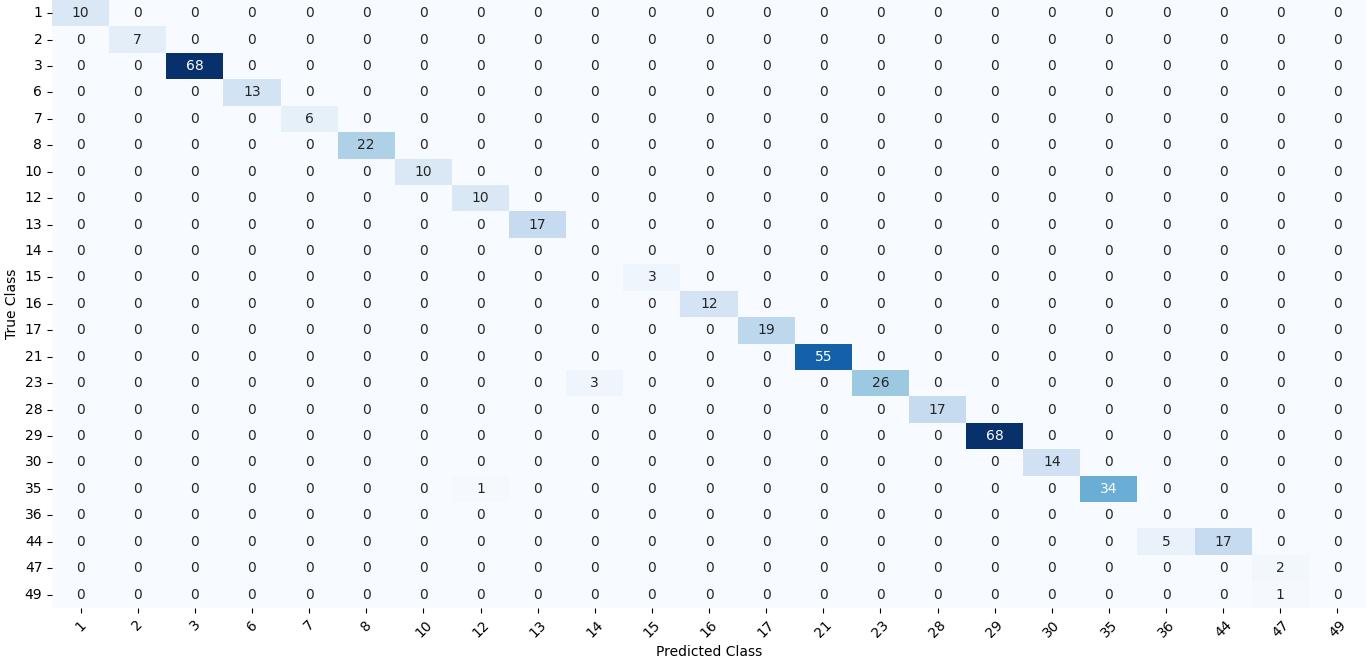}
\caption{Confusion matrix for 50 classes}
\label{fig_110}
\end{figure}

The number of clusters affects the efficiency of the CQI mapping operation, so it has been tested with different cluster counts. Figures \ref{fig_17} to \ref{fig_110} display the confusion matrices from the test process for 20, 30, 40, and 50 clusters, respectively.

\begin{figure}[!t]
\centering
\includegraphics[width=3.4in]{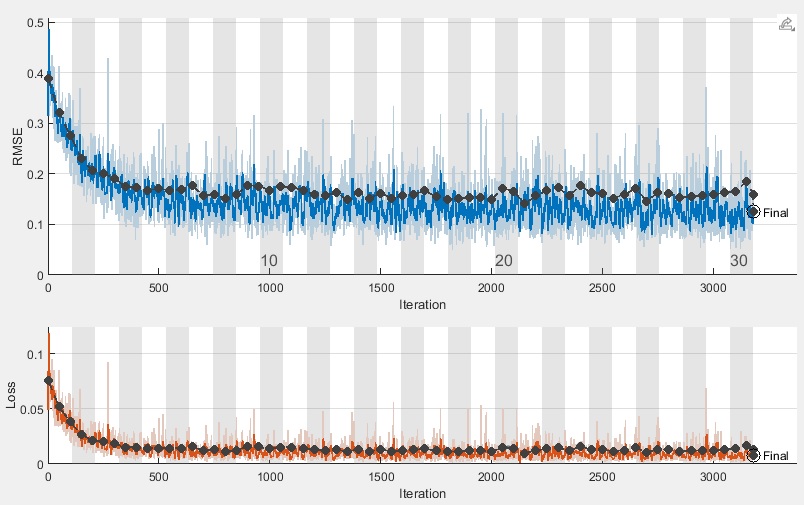}
\caption{Training process of the block error rate estimation model}
\label{fig_111}
\end{figure}

As shown in Figures \ref{fig_17} to \ref{fig_110}, the testing process demonstrates acceptable efficiency between different cluster numbers, with the 40-cluster case performing the best. Therefore, the BLER estimation model is based on 40 clusters. Figure \ref{fig_111} illustrates the DNN's training process for BLER estimation.

\subsection{Comparison of the proposed algorithm's performance with benchmark algorithms}
\begin{figure}[!t]
\centering
\includegraphics[width=3.4in]{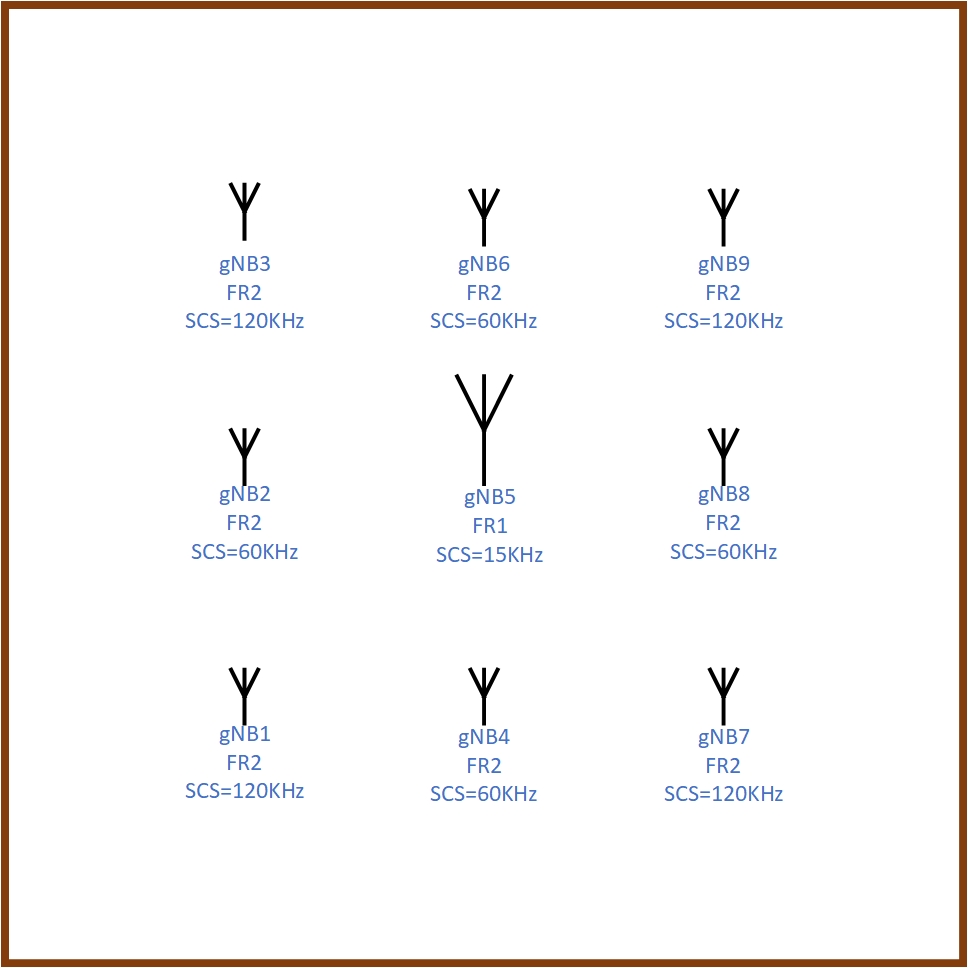}
\caption{Geographical area of simulation}
\label{fig_112}
\end{figure}

To evaluate the performance of the proposed algorithm, it is compared with two benchmark algorithms: the SNR-based single-connectivity algorithm and the DRL-based method presented in \cite{hernandez2022deep1}, referred to as DRLMC. The SNR-based method selects the closest gNB to the UE, while the DRLMC algorithm uses a DRL approach to select the serving cluster for the UE. For comparison, a simulation scenario was created with nine gNBs positioned in a 1000-meter square area. A gNB operating in FR1 with OFDM numerology 0 is placed at the center, while the other gNBs operate in FR2 with numerologies 2 or 3. Figure \ref{fig_112} shows the simulation environment. Each gNB has 66 resource blocks in the PDSCH, with resource availability fluctuating over time due to network dynamics. The selection of the UE serving cluster is handled by the central unit, and the downlink traffic is allocated proportionally among the members of the serving cluster. Table \ref{tab:table3} outlines the simulation settings.

\begin{table}[!t]
\caption{Simulation Setup\label{tab:table3}}
\centering
\begin{tabular}{|c||c|}
\hline
\textbf{Parameter} & \textbf{Value}\\
\hline
Geographical Area & 1000 $\times$ 1000 meters\\
\hline
Number of gNBs  & 9 \\
\hline
Sub-carrier Spacing & 15, 60, 120 KHz \\
\hline
Number of RBs in each gNB & 66 \\
\hline
Macro-cell Frequency Range & sub-7 GHz \\
\hline
Small Cell Frequency Range & 25 GHz \\
\hline
Channel Model & CDL \\
\hline
Macro-cell Maximum Transmission Power & 49 dBm \\
\hline
Small Cell Maximum Transmission Power & 29 dBm \\
\hline
Simulation Time & 600 seconds \\
\hline
UE Movement Model & Random waypoint \\
\hline
\end{tabular}
\end{table}
To evaluate the efficiency of the proposed method against benchmark algorithms, a UE is defined with QoS requirements of 140 Mbps for data rate, 0.4 milliseconds for latency, and 0.99 for reliability. In this evaluation, the coefficients in Equation \ref{eq16} are set equally for QoS requirements and spectrum efficiency in the proposed algorithm. Table \ref{tab:table4} summarizes the application features used in this section.

\begin{table}[!t]
\caption{QoS Characteristics of UE\label{tab:table4}}
\centering
\begin{tabular}{|c||c|}
\hline
\textbf{Parameter} & \textbf{Value}\\
\hline
Rate Requirement & 150 Mbps\\
\hline
Reliability Requirement & 0.99 \\
\hline
Latency Requirement & 0.4 ms \\
\hline
Impact Factor of Rate & 0.25 \\
\hline
Impact Factor of Reliability & 0.25 \\
\hline
Impact Factor of Latency & 0.25 \\
\hline
Impact Factor of Spectrum Efficiency & 0.25 \\
\hline
Packet Size & 1500 Bytes \\
\hline
\end{tabular}
\end{table}
The measured performance metrics are as follows:
\begin{itemize}
    \item \textbf{Average data rate:} Defined as the ratio of the received data size to the duration of the data transfer
    \item \textbf{Average latency:} The time elapsed from when the packet is placed on the channel to its successful decoding at the receiver
    \item \textbf{Reliability:} Defined as the ratio of received packets to the total number of sent packets, including HARQ re-transmissions. A packet is considered not received if errors are not corrected by HARQ retransmissions.
    \item \textbf{Average resource consumption:} The average number of RBs used, multiplied by the sub-carrier spacing and 12. Each RB consists of 12 sub-carriers, with the sub-carrier spacing varying based on the configuration of the selected gNBs.
    \item \textbf{Spectrum efficiency:} Defined as the ratio of the average data rate to the average resource consumption over a given period of time.
\end{itemize}
Since the serving cluster for the UE includes multiple gNBs, the data rate and resource consumption are determined by summing the experienced rates and consumed resources of each cluster member. For error rate and average latency, the weighted average of the cluster members' error rates and latencies is used, with the weight for each gNB being the rate from that gNB divided by the total rate received. 

\begin{figure}[!t]
\centering
\includegraphics[width=3.2in]{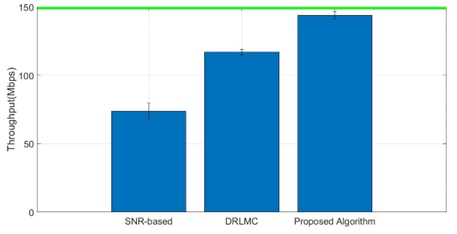}
\caption{UE's average data rate in the proposed algorithm along with the benchmark algorithms}
\label{fig_113}
\end{figure}

\begin{figure}[!t]
\centering
\includegraphics[width=3.2in]{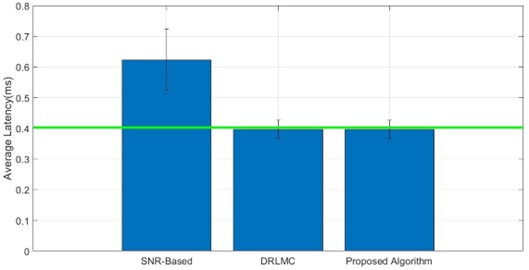}
\caption{UE's average latency in the proposed algorithm along with the benchmark algorithms}
\label{fig_114}
\end{figure}

\begin{figure}[!t]
\centering
\includegraphics[width=3.2in]{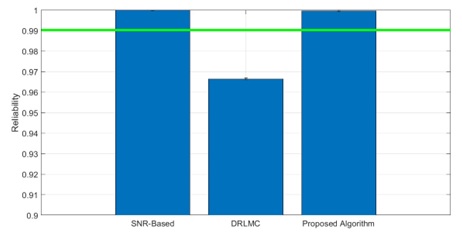}
\caption{UE's average reliability in the proposed algorithm along with the benchmark algorithms}
\label{fig_115}
\end{figure}

\begin{figure}[!t]
\centering
\includegraphics[width=3.2in]{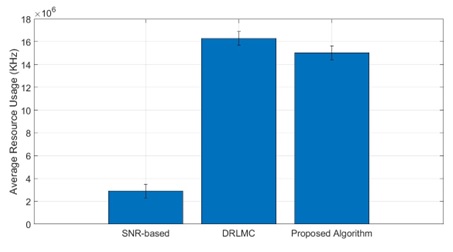}
\caption{Average resource usage in the proposed algorithm along with the benchmark algorithms}
\label{fig_116}
\end{figure}

\begin{figure}[!t]
\centering
\includegraphics[width=3.3in]{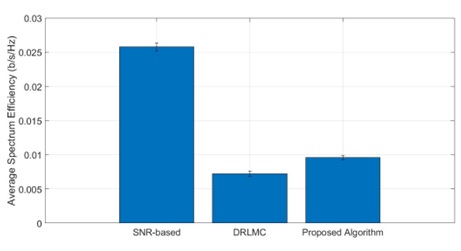}
\caption{Average spectrum efficiency in the proposed algorithm along with the benchmark algorithms}
\label{fig_117}
\end{figure}

Figures \ref{fig_113} to \ref{fig_117} present the simulation results. To ensure reliability, the results are based on 10 simulation repetitions, each with a different random number generator seed. The reported results show the average and standard deviation of the values obtained.

Figure \ref{fig_113} illustrates the average data rate achieved by the proposed algorithm compared to the benchmark algorithms.The proposed algorithm provides a higher data rate, closer to the UE's requested rate, compared to the benchmark algorithms. The difference in data rates between the SNR-based algorithm and the other two is due to the use of multi-connectivity in both the DRLMC and the proposed algorithm. Additionally, the proposed algorithm achieves a higher data rate than the DRLMC. This can be attributed to the distinct approaches these two algorithms take in selecting the UE's serving cluster. While DRLMC focuses on load balancing, which can impact the QoS of UEs, the proposed algorithm prioritizes the UE's QoS requirements.

Figure \ref{fig_114} shows the average experienced latency of the proposed algorithm compared to the benchmark algorithms. The SNR-based algorithm results in higher latency than the other two, primarily due to its focus on selecting the nearest gNB without considering latency. This often leads to the selection of gNBs with shorter sub-carrier spacing, causing longer time slots. In contrast, the DRLMC algorithm, which prioritizes the UE's data rate requirement, selects gNBs with higher sub-carrier spacing that can provide higher data rates, thus achieving lower latency than the SNR-based algorithm. Similarly, the proposed algorithm also shows lower latency, as it considers both the data rate and latency in the cluster selection process.

Figure \ref{fig_115} shows the average reliability of the proposed algorithm compared to the benchmark algorithms. The SNR-based algorithm offers higher reliability than the other two algorithms, due to its selection of the nearest gNB, which provides better channel conditions. The DRLMC algorithm, on the other hand, results in lower reliability than expected because it does not prioritize reliability in its decision-making process. In contrast, the proposed algorithm successfully meets the required reliability threshold by explicitly considering the application's reliability requirements during the serving cluster selection process.

Figure \ref{fig_116} depicts the average resource consumption. The SNR-based method consumes the least resources due to its single-connectivity approach, while DRLMC consumes more resources because it focuses on load balancing. The proposed algorithm consumes fewer resources than DRLMC, as it optimizes resource usage while meeting QoS requirements.

Figure \ref{fig_117} shows the spectrum efficiency of the proposed algorithm alongside the benchmark algorithms. As expected, the SNR-based method shows higher spectrum efficiency due to better channel conditions with the nearest gNB and its use of higher-order modulation and coding schemes. It also consumes fewer resources by using a single connection, further improving its spectrum efficiency. However, the proposed algorithm outperforms the DRLMC method in spectrum efficiency by achieving a higher data rate while using fewer resources.

\begin{figure}[!t]
\centering
\includegraphics[width=3.2in]{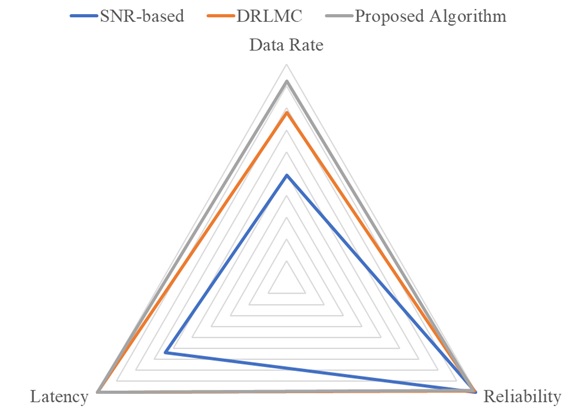}
\caption{How the proposed algorithm and benchmark algorithms pay attention to the QoS of UE}
\label{fig_118}
\end{figure}

Figure \ref{fig_118} illustrates the QoS provisioning of the proposed algorithm compared to benchmark algorithms. A score of 1 is assigned when the provided QoS metric exceeds the application's required level. If the provided QoS metric is lower than required, the score approaches 0. As shown in Figure \ref{fig_118}, the proposed algorithm successfully balances the QoS requirements of the application.

\section{Conclusion}
This paper proposes a QoS-aware multi-connectivity framework that integrates machine learning techniques to address the challenges of modern network environments in meeting diverse QoS requirements. By leveraging Deep Neural Networks, the algorithm estimates the block error rate of various UE-BS connections, enhancing serving cluster selection and optimizing data rate splitting among connected gNBs, with the primary goal of fulfilling UE QoS requirements. Performance evaluation results demonstrate that the proposed approach significantly improves data rate, reliability, and latency, outperforming standard methods and contemporary algorithms in critical scenarios. Additionally, the enhanced spectrum efficiency offers a notable advantage, paving the way for broader applications in next-generation communication systems. 
For future work, several key areas can be explored: refining the algorithm's adaptability to diverse network conditions and user behavior patterns, integrating advanced machine learning techniques such as reinforcement learning to enhance real-time decision-making, and extending the framework to support heterogeneous networks by incorporating a wider range of communication technologies.

\section{}

\newpage

\section{Biography Section}
 
\vspace{11pt}

\vspace{-33pt}
\begin{IEEEbiography}[{\includegraphics[width=1in,height=1.25in,clip,keepaspectratio]{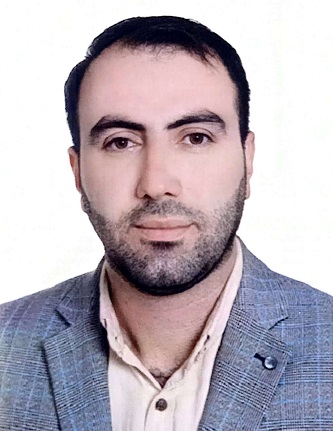}}]{Ali Parsa}
received the B.S. from Ilam University, Ilam, Iran in 2013 and M.S. and Ph.D. from University of Isfahan, Isfahan, Iran in 2017 and 2024 respectively. His research interests are $5G\&B$ networks, machine learning usage in computer networks, and quality of service.
\end{IEEEbiography}

\vspace{11pt}

\vspace{-33pt}
\begin{IEEEbiography}[{\includegraphics[width=1in,height=1.25in,clip,keepaspectratio]{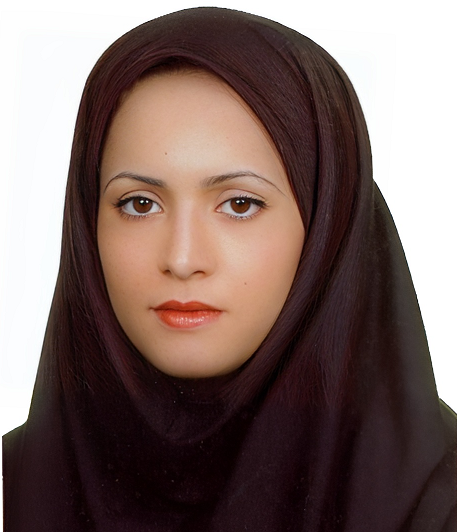}}]{Neda Moghim}
received her B.Sc. and 
M.Sc. degrees from Isfahan University 
of Technology, Isfahan, Iran, in 1999 
and 2002, respectively, and her Ph.D. 
from Amirkabir University of 
Technology, Tehran, Iran, in 2009. She 
is an Associate Professor in the 
Department of Computer Engineering 
at the University of Isfahan, Iran, and a 
Research Assistant Professor at the Center for Secure and 
Intelligent Critical Systems (CSICS), Office of Enterprise 
Research and Innovation (OERI), and the School of 
Cybersecurity at Old Dominion University (ODU), Suffolk, 
VA, USA. She has authored numerous technical papers in 
telecommunications journals and conferences. Her research
interests include $5G\&B$ Networks, the Internet of Things, and Semantic Communication.
\end{IEEEbiography}

\begin{IEEEbiography}[{\includegraphics[width=1in,height=1.25in,clip,keepaspectratio]{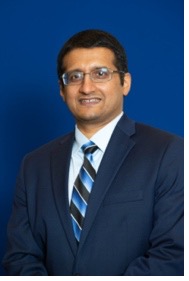}}]{Sachin Shetty}
received the Ph.D. degree in modeling and simulation from Old Dominion University in 2007. He was an Associate Professor with the Electrical and Computer Engineering Department, Tennessee State University, USA. He is currently a Executive Director with the Center for Secure and Intelligent Critical Systems and Professor in the Department of Electrical and Computer Engineering.He has authored or coauthored over 200 research articles in journals and conference proceedings and two books. His research interests lie at the intersection of computer networking, network security, and machine learning.

\end{IEEEbiography}

\vfill

\end{document}